\documentclass[aps,prd,preprintnumbers,superscriptaddress,nofootinbib,twocolumn]{revtex4-2}

\usepackage{amsmath,amssymb,amsfonts,bm}
\usepackage{graphicx}
\usepackage{xcolor}
\definecolor{newrev}{RGB}{0,120,120}
\definecolor{finalrev}{RGB}{220,100,0}
\definecolor{latestrev}{RGB}{120,40,170}
\usepackage{hyperref}
\usepackage{booktabs}
\usepackage{microtype}

\hypersetup{colorlinks=true,citecolor=blue,linkcolor=blue,urlcolor=blue}

\newcommand{\st}[1]{\text{\tiny\rm #1}}
\newcommand{\variety}{V}
\newcommand{\icm}{I_{\st{cm}}}
\newcommand{\vnew}{V_{\st{New}}}
\newcommand{\Etot}{E_{\st{tot}}}
\newcommand{\Jtot}{\mathbf J_{\st{tot}}}
\newcommand{\Ptot}{\mathbf P_{\st{tot}}}
\newcommand{\Aani}{\mathcal A_6}

\begin{document}

\title{Structural morphology and the gravitational arrow of time}

\author{Julian Barbour}
\email{julian.barbour@physics.ox.ac.uk}
\affiliation{College Farm, The Town, South Newington, Banbury, OX15 4JG, United Kingdom}

\author{Francisco S. N. Lobo}
\email{fslobo@ciencias.ulisboa.pt}
\affiliation{Instituto de Astrof\'isica e Ci\^encias do Espa\c{c}o,
	Faculdade de Ci\^encias da Universidade de Lisboa, Edif\'icio C8,
	Campo Grande, P-1749-016 Lisbon, Portugal}
\affiliation{Departamento de F\'isica, Faculdade de Ci\^encias da Universidade de Lisboa,
	Edif\'icio C8, Campo Grande, P-1749-016 Lisbon, Portugal}

\author{Maria I. R. Louren\c{c}o}
\email{fc56407@alunos.fc.ul.pt}
\affiliation{Instituto de Astrof\'isica e Ci\^encias do Espa\c{c}o,
	Faculdade de Ci\^encias da Universidade de Lisboa, Edif\'icio C8,
	Campo Grande, P-1749-016 Lisbon, Portugal}

\date{\today}

\begin{abstract}
	The Newtonian $N$-body problem in the zero-energy, zero-linear-momentum,
	and zero-angular-momentum sector provides a time-reversal-invariant
	setting in which generic complete solutions possess a Janus point and
	exhibit a gravitational arrow on both branches away from it. The
	dimensionless variety $\variety$, a global scale-invariant measure of
	clustering contrast, is not pointwise monotonic but fluctuates while
	growing between rising bounds away from the Janus region. Central
	configurations are critical shapes of the same scale-invariant
	landscape and therefore provide controlled probes of the structural
	information encoded by $\variety$.
	We investigate this question for two planar $N=5000$ central
	configurations using the numerical particle-coordinate data.
	Local morphology is quantified by the six-neighbour anisotropy $A_6$,
	which ranges from $0$ for an isotropic local environment to $1$ for an
	effectively one-dimensional one. Although the varieties of the two
	configurations differ by only $1.686\%$, their mean anisotropies differ
	by $156.8\%$, from $0.1331$ to $0.3419$. Moreover, $18.1\%$ of the
	particles in the higher-variety configuration satisfy $A_6>0.5$,
	whereas none do so in the lower-variety configuration. The anisotropy
	ordering persists for all tested neighbourhood sizes $4\le k\le12$,
	and the principal contrast survives a dimensionless close-pair
	robustness test.
	For this pair of critical shapes, nearby values of the global variety
	therefore coexist with markedly different local geometrical
	organization. Thus the scalar quantity whose long-term behavior
	characterizes the BKM gravitational arrow does not, by itself, uniquely
	specify morphology. This identifies local relational observables as a
	complementary level of description and provides a quantitative bridge
	between the static shape-space landscape and morphology along genuine
	Janus-point histories.
\end{abstract}

\maketitle

\section{Introduction}

The origin of an arrow of time in systems governed by
time-reversal-invariant microscopic laws is conventionally associated
with special boundary conditions. In ordinary statistical mechanics,
this idea is embodied in the low-entropy Past Hypothesis.
Self-gravitating systems are qualitatively different: gravity is
long ranged, clustering is dynamically favored, and isolated
self-gravitating systems can exhibit behavior, such as negative heat
capacities, that differs fundamentally from that of ordinary extensive
thermodynamic systems. These features motivate a complementary
question: can an intrinsic arrow arise from gravitational dynamics
itself, without imposing a special low-variety boundary condition at
one temporal endpoint?

Barbour, Koslowski, and Mercati (BKM) provided a particularly clean
Newtonian realization of such an intrinsic gravitational arrow
\cite{Barbour:2014bga,Barbour:2013jya}. Consider the $N$-body problem in the
sector
\begin{equation}
	\Etot=0,\qquad
	\Ptot=0,\qquad
	\Jtot=0 .
	\label{eq:sector}
\end{equation}
For the generic complete two-sided solutions of interest here, the
center-of-mass moment of inertia possesses a unique minimum, the
\emph{Janus point}. In the BKM formulation, the dimensionless variety
$\variety$ is not pointwise monotonic; rather, it fluctuates while
growing between rising bounds away from the Janus region
\cite{Barbour:2014bga,Barbour:2013jya}. On both branches, the long-time
evolution is accompanied by the formation of bound subsystems and
records.

A distinct line of work has explored large-$N$ \emph{central
	configurations} (CCs), which are critical shapes of the normalized
Newtonian potential, equivalently of the scale-invariant variety for
fixed masses. Numerical searches reveal highly uniform configurations
near the lowest values of variety and, at only modestly larger values
of $\variety$, configurations displaying filaments, loops, and
void-like regions
\cite{Battye2003,Lourenco:2026uto,Lourenco:2026lbr}. This observation is
suggestive, but two logically distinct notions must be kept separate.
The Janus point is defined by the minimum of the overall scale variable
$\icm$ and is not, in general, a critical point or minimum of
$\variety$. Central configurations, by contrast, are distinguished
critical shapes of the scale-invariant variety; generic Newtonian
histories are not required to pass through them. The dynamical BKM
arrow and the static CC landscape must therefore not be identified.

The purpose of the present work is to place these two ingredients in a
common quantitative framework. We use the numerical
particle-coordinate data for two previously generated planar $N=5000$
CCs obtained in the numerical analysis reported in the revised version
of Ref.~\cite{Lourenco:2026uto} and apply the same local morphology
diagnostic to both. Their varieties differ by only $1.686\%$, whereas
their mean six-neighbour anisotropies differ by $156.8\%$. The result
therefore establishes a large quantitative morphological contrast
between two critical shapes of shape space.

This comparison is relevant to the gravitational arrow because variety
is a single global, scale-invariant measure of clustering contrast,
whereas structures such as filaments, loops, and void-like regions
involve local geometrical organization. Determining whether nearby
values of variety can support markedly different morphologies therefore
tests how completely this global relational scalar specifies the
structural content of a configuration. For the two critical shapes
studied here, our results show that variety alone does not uniquely
specify local morphology. This indicates that additional relational
observables may be needed for a more complete quantitative description
of structural organization, without diminishing the role of variety
in characterizing the BKM gravitational arrow.

The local anisotropy introduced below provides one such complementary
observable. Importantly, the same diagnostic can in principle be
evaluated both on central configurations and along genuine Newtonian
trajectories. It therefore provides a common quantitative language with
which the morphology of the static CC landscape can subsequently be
compared with that sampled dynamically along Janus-point histories.
The present two-configuration comparison does not itself establish such
a dynamical relation; it identifies and quantifies the structural
distinction that motivates that test.

The paper is organized to keep the dynamical and static ingredients
logically separate before comparing them. Section~\ref{sec:relational}
reviews the relational Newtonian dynamics and the Janus-point
construction. Section~\ref{sec:variety} defines $\variety$ and summarizes
its long-term behavior in the BKM framework. Section~\ref{sec:CC}
introduces central configurations as critical probes of shape space and
specifies the two planar $N=5000$ configurations and numerical
particle-coordinate data used in the present analysis.
Section~\ref{sec:morphology} introduces the local morphology observable
and presents the numerical results and robustness tests.
Section~\ref{sec:bridge} then compares the static shape-space landscape
with the dynamical gravitational arrow and formulates the complementary
static-ensemble and Janus-trajectory tests. Finally,
Sec.~\ref{sec:discussion} discusses the physical implications and
summarizes the main conclusions.

\section{Relational Newtonian dynamics}
\label{sec:relational}

Let $m_a$, $\bm r_a$, and $\bm p^a$ denote the masses, positions, and
canonical momenta of $N$ point particles. We set $G=1$. The Newtonian
potential is
\begin{equation}
	\vnew=-\sum_{a<b}\frac{m_am_b}{r_{ab}},
	\qquad
	r_{ab}=|\bm r_a-\bm r_b|,
	\label{eq:Vnew}
\end{equation}
and the total energy is
\begin{equation}
	\Etot=\sum_a\frac{\bm p^a\cdot\bm p^a}{2m_a}+\vnew .
	\label{eq:energy}
\end{equation}
The total linear and angular momenta are
\begin{equation}
	\Ptot=\sum_a\bm p^a,\qquad
	\Jtot=\sum_a\bm r_a\times\bm p^a .
\end{equation}
We work in the sector defined by Eq.~\eqref{eq:sector}. Since
$\Ptot=0$, the center of mass moves uniformly and we choose the
center-of-mass frame, with
\begin{equation}
	\bm r_{\rm cm}=\frac{1}{m_{\rm tot}}\sum_a m_a\bm r_a,
	\qquad
	m_{\rm tot}=\sum_a m_a .
\end{equation}

The center-of-mass moment of inertia is
\begin{equation}
	\icm=\sum_a m_a|\bm r_a-\bm r_{\rm cm}|^2
	=\frac{1}{m_{\rm tot}}\sum_{a<b}m_am_b r_{ab}^2 .
	\label{eq:Icm}
\end{equation}
For a potential homogeneous of degree $-1$, the Lagrange--Jacobi
identity gives
\begin{equation}
	\ddot{I}_{\st{cm}}
	=4\Etot-2\vnew .
	\label{eq:LJgeneral}
\end{equation}
Hence, in the zero-energy sector,
\begin{equation}
	\ddot{I}_{\st{cm}}=-2\vnew>0 .
	\label{eq:LJ}
\end{equation}
Since $\vnew<0$ at every finite collision-free configuration,
$\icm(t)$ is strictly convex along every collision-free interval of a
solution.

The dilatational momentum is
\begin{equation}
	D=\sum_a(\bm r_a-\bm r_{\rm cm})\cdot\bm p^a
	=\frac12\dot{I}_{\st{cm}} .
	\label{eq:D}
\end{equation}
Equation~\eqref{eq:LJ} then yields
\begin{equation}
	\dot D=-\vnew>0 .
	\label{eq:Dmono}
\end{equation}
Thus $D$ is strictly increasing and can cross zero at most once.
For the generic complete two-sided collision-free zero-energy histories
considered here, $D$ takes both negative and positive values; it
therefore has a unique zero $t_J$,
\begin{equation}
	D(t_J)=0,\qquad
	\icm(t_J)=\min_t\icm(t) .
	\label{eq:Janus}
\end{equation}
This unique minimum of $\icm$ defines the Janus point. The distinction
important for what follows is simply that this is a statement about the
scale of the configuration: it does not require the scale-invariant
variety $\variety$ to be stationary or minimal at $t_J$.

Newtonian gravity also possesses the dynamical similarity
\begin{equation}
	\bm r_a(t)\longmapsto
	\alpha\,\bm r_a\!\left(t/\alpha^{3/2}\right),
	\qquad \alpha>0 ,
	\label{eq:similarity}
\end{equation}
which changes the overall dimensional scale while preserving the
corresponding unparametrized curve in shape space. This motivates
describing the intrinsic dynamics after quotienting translations,
rotations, and global dilatations.

\section{Variety and the gravitational arrow}
\label{sec:variety}

A convenient dimensionless measure of global clustering contrast is the
\emph{variety} $\variety$ (the BKM shape complexity), defined by
\begin{equation}
	\variety=\frac{\ell_{\rm rms}}{\ell_{\rm mhl}},
	\label{eq:CsDef}
\end{equation}
where
\begin{align}
	\ell_{\rm rms}
	&=
	\frac{1}{m_{\rm tot}}
	\sqrt{\sum_{a<b}m_am_b r_{ab}^{2}}
	=
	\sqrt{\frac{\icm}{m_{\rm tot}}},
	\label{eq:lrms}\\
	\ell_{\rm mhl}^{-1}
	&=
	\frac{1}{m_{\rm tot}^{2}}
	\sum_{a<b}\frac{m_am_b}{r_{ab}}
	=
	-\frac{\vnew}{m_{\rm tot}^{2}} .
	\label{eq:lmhl}
\end{align}
It follows that
\begin{equation}
	\variety
	=
	-\frac{\sqrt{\icm}\,\vnew}{m_{\rm tot}^{5/2}} .
	\label{eq:Cs}
\end{equation}
Because $\sqrt{\icm}$ scales linearly with length whereas $\vnew$
scales inversely with length, $\variety$ is invariant under a global
dilatation. Since it is constructed entirely from pair separations, it
is also invariant under translations and rotations. For fixed masses,
$\variety$ therefore depends only on the relational shape of the
configuration and defines a dimensionless function on shape space.

The two length scales entering Eq.~\eqref{eq:CsDef} emphasize different
parts of the distribution: $\ell_{\rm rms}$ is especially sensitive to
large pair separations, whereas $\ell_{\rm mhl}$ is strongly affected by
small separations. Their ratio therefore compares the overall extent of
the configuration with its short-distance clustering. A system
containing tight subsystems within a much larger distribution can
accordingly have a large value of $\variety$. Variety is thus a
geometrical measure of clustering contrast, not a thermodynamic entropy.

The asymptotic organization of Newtonian $N$-body solutions is
constrained by the Marchal--Saari results \cite{MarchalSaari}, which
describe their long-time decomposition into bound and mutually
receding subsystems. In the BKM framework, this asymptotic structure is
accompanied by a characteristic behavior of the scale-invariant
variety: $\variety$ is not pointwise monotonic but fluctuates while
growing between rising bounds on each branch away from the Janus region
\cite{Barbour:2014bga,Barbour:2013jya}. The gravitational arrow considered here
therefore refers to this long-term branch-wise behavior of clustering
contrast, not to a strictly monotonic increase of $\variety$ at every
instant.

As emphasized in Sec.~\ref{sec:relational}, the Janus point itself is
defined by the minimum of the scale variable $\icm$; its definition
does not require $\variety$ to be stationary or minimal there. No
microscopic dissipation is implied, and no identification of $\variety$
with thermodynamic entropy is assumed.

For the present work, the essential point is that $\variety$ compresses
the global relational information contained in all pair separations
into a single scalar. Whether nearby values of this scalar can
nevertheless correspond to markedly different local geometrical
organization is a separate question. The central configurations
considered next provide controlled critical shapes with which to
investigate precisely this issue.

\section{Central configurations and the numerical shape-space landscape}
\label{sec:CC}

A central configuration (CC) is a configuration for which the
Newtonian acceleration of every particle is proportional to its
displacement from the center of mass. It satisfies
\begin{equation}
	\sum_{b\ne a}m_b
	\frac{\bm r_b-\bm r_a}{r_{ab}^3}
	=
	-\Lambda(\bm r_a-\bm r_{\rm cm}),
	\qquad \Lambda>0 .
	\label{eq:CC}
\end{equation}
Equivalently, CCs are stationary points of $\vnew$ subject to fixed
$\icm$ and fixed center of mass \cite{Saari,MeyerHallOffin}. Since, for
fixed masses,
\begin{equation}
	\variety=-\frac{\sqrt{\icm}\,\vnew}{m_{\rm tot}^{5/2}},
\end{equation}
the corresponding shapes are critical points of the scale-invariant
variety on shape space. With appropriate initial velocities, central
configurations underlie the familiar homothetic and homographic
solutions of the Newtonian $N$-body problem.

Their role here is variational: they provide distinguished critical
shapes with which to probe the structure of the variety landscape.
They should not be interpreted as successive configurations along a
generic Newtonian history. Large-$N$ numerical searches have shown that
such critical shapes can display remarkably diverse morphologies
\cite{Battye2003,Lourenco:2026uto,Lourenco:2026lbr}. Low-$\variety$
planar configurations can exhibit highly regular, nearly isotropic
local organization, whereas configurations at somewhat larger
$\variety$ can display filamentary networks, curved structures, loops,
and void-like domains. This motivates the quantitative question
considered here: how different can the local morphology of critical
shapes be when their global varieties are close?

\begin{figure*}[t]
	\centering
	\includegraphics[width=0.45\textwidth]{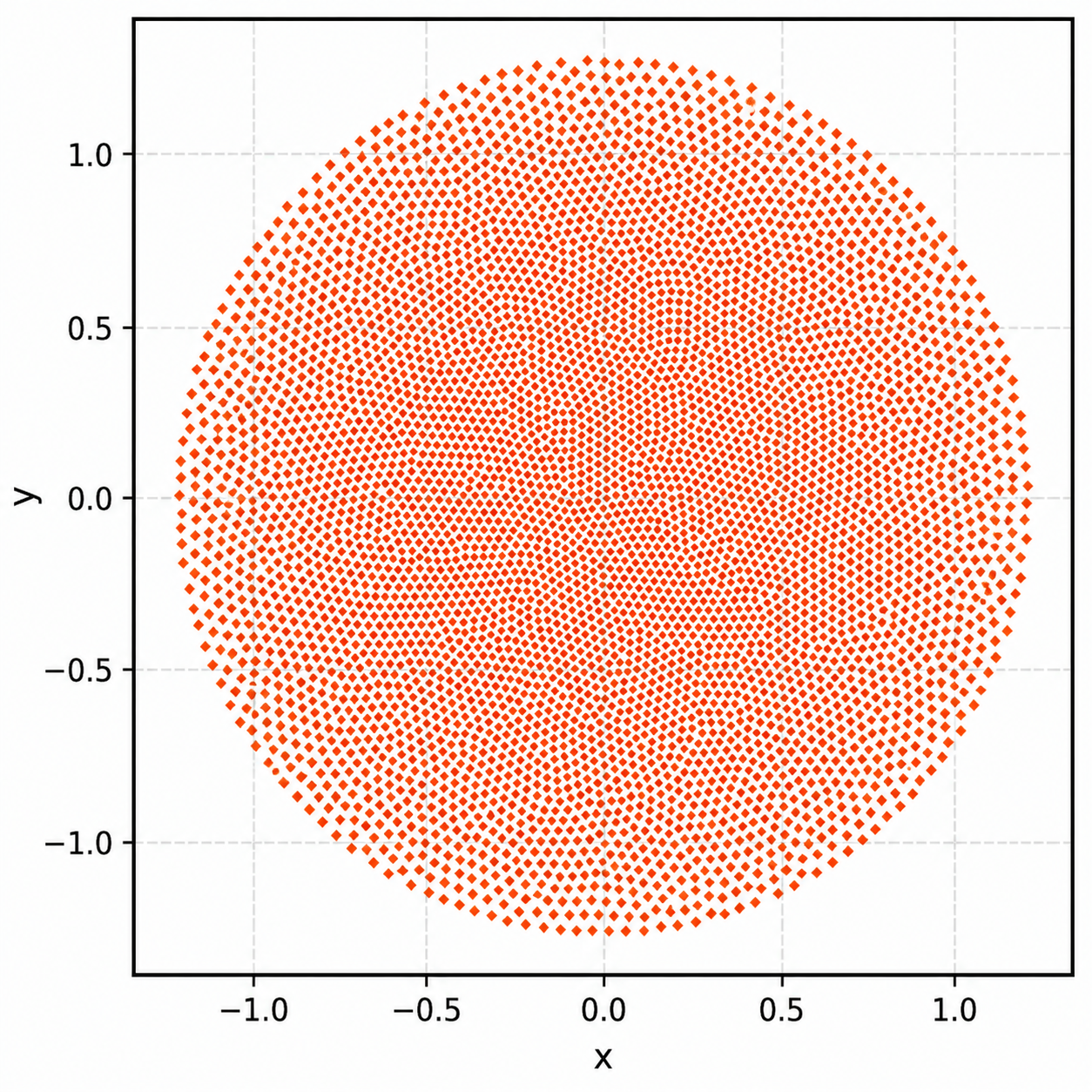}
	\hfill
	\includegraphics[width=0.45\textwidth]{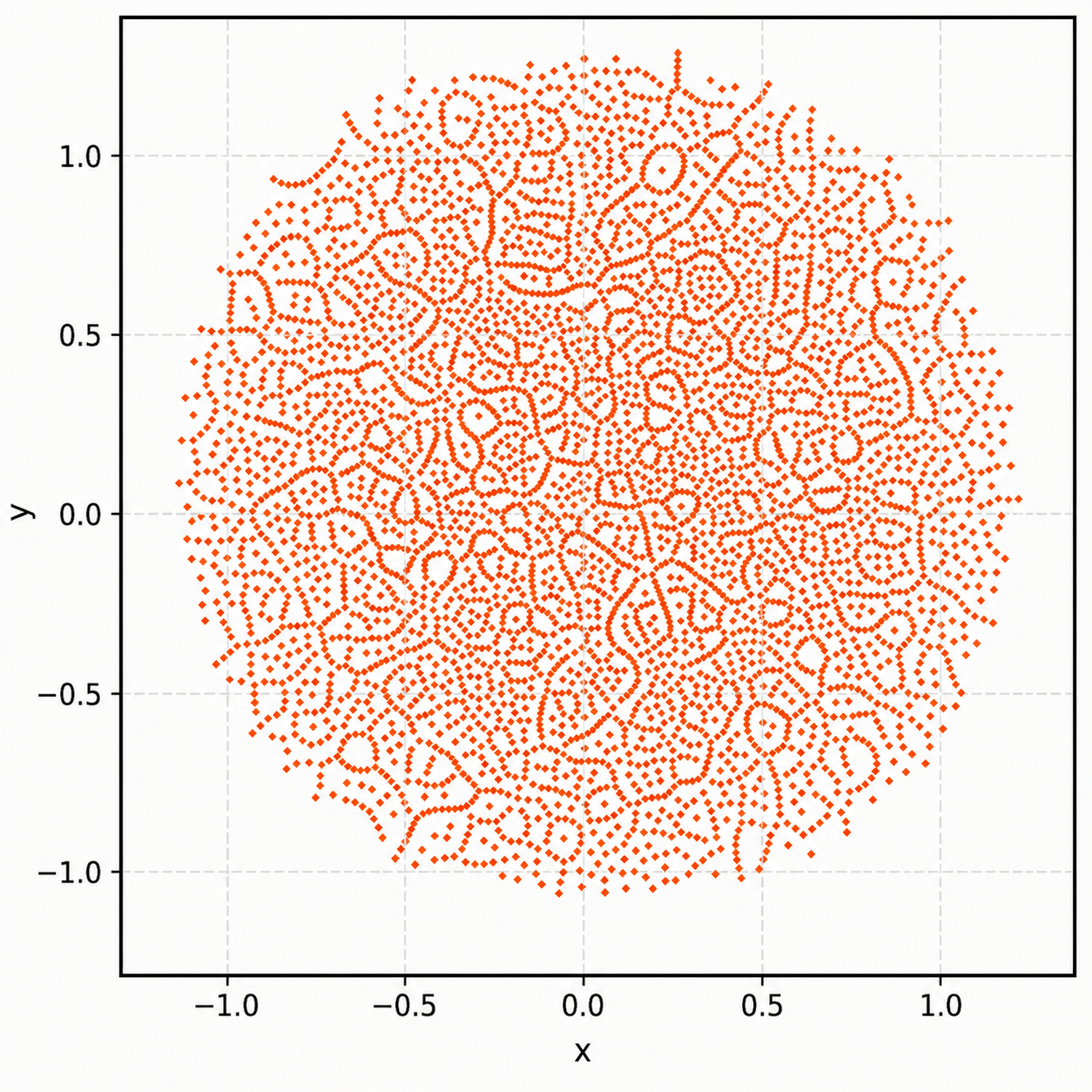}
	\caption{Planar central configurations of $N=5000$ equal-mass
		particles analyzed in this work, with varieties
		$\variety=0.585701$ (left) and $\variety=0.595574$ (right).
		The numerical particle-coordinate data underlying these
		configurations, obtained in the numerical analysis reported in the
		revised version of Ref.~\cite{Lourenco:2026uto}, are used directly in
		the present analysis. Despite their $1.686\%$ relative difference in
		variety, the two configurations exhibit markedly different local
		organization.}
	\label{fig:planarCCs}
\end{figure*}

We examine two planar equal-mass $N=5000$ central configurations
obtained in the numerical analysis reported in the revised version of
Ref.~\cite{Lourenco:2026uto}, using directly the numerical
particle-coordinate data underlying them. Their varieties are
\begin{equation}
	\variety^{(0)}=0.585701,\qquad
	\variety^{(1)}=0.595574 ,
	\label{eq:Cvalues}
\end{equation}
with relative separation
\begin{equation}
	\frac{\variety^{(1)}-\variety^{(0)}}{\variety^{(0)}}
	=1.686\% .
	\label{eq:Cdifference}
\end{equation}
The first is the lowest-variety $N=5000$ planar configuration obtained
in that numerical search. Despite the small difference in variety, the
second exhibits pronounced directional structure, including
filamentary strands, curved structures, loops, and void-like regions.
The two configurations are displayed in Fig.~\ref{fig:planarCCs}.

For the equal-mass configurations considered here, $m_a=1$ and
$m_{\rm tot}=N$. Equations~\eqref{eq:lrms}--\eqref{eq:lmhl} reduce to
\begin{align}
	\ell_{\rm rms}
	&=
	\frac{1}{N}
	\left(\sum_{a<b}r_{ab}^{2}\right)^{1/2},
	\label{eq:lrmsEqual}\\
	\ell_{\rm mhl}^{-1}
	&=
	\frac{1}{N^{2}}
	\sum_{a<b}\frac{1}{r_{ab}},
	\label{eq:lmhlEqual}
\end{align}
and hence
\begin{equation}
	\variety=
	\frac{1}{N^3}
	\left(\sum_{a<b}r_{ab}^{2}\right)^{1/2}
	\left(\sum_{a<b}\frac{1}{r_{ab}}\right).
	\label{eq:CsEqual}
\end{equation}
The values in Eq.~\eqref{eq:Cvalues} were evaluated from the 
numerical particle-coordinate data using Eq.~\eqref{eq:CsEqual}.

The visual contrast in Fig.~\ref{fig:planarCCs} motivates an independent
quantitative characterization of local morphology. This is important
because $\variety$ and the morphology observable introduced below probe
different information: $\variety$ is determined by all pair
separations, whereas the local anisotropy statistic measures the
directional organization of neighbouring particles. Although both are
evaluated from the same coordinates, the latter is neither defined
from nor fitted to $\variety$.

The question addressed by the present comparison is therefore strictly
static: whether two critical shapes with nearby global varieties can
possess markedly different local organization. Whether any analogous
morphology--variety relation is statistically realized along genuine
Janus-point histories is a separate dynamical question considered in
Sec.~\ref{sec:bridge}.

\section{A common scale-invariant morphology observable}
\label{sec:morphology}

The visual contrast in Fig.~\ref{fig:planarCCs} motivates an independent
quantitative measure of local organization. We require a diagnostic that
is applied identically to both configurations, is invariant under
translations, rotations, and global dilatations, and distinguishes
locally isotropic environments from directionally organized ones.

For each particle $a$, let ${\cal N}_k(a)$ denote its $k$ nearest
neighbours. Define the $2\times2$ local second-moment tensor
\begin{equation}
	Q^{(a,k)}_{ij}
	=
	\frac{1}{k}\sum_{b\in{\cal N}_k(a)}
	\Delta r^{(ab)}_i\Delta r^{(ab)}_j,
	\qquad
	\Delta\bm r^{(ab)}=\bm r_b-\bm r_a .
	\label{eq:Qk}
\end{equation}
Let $q_1^{(a,k)}\ge q_2^{(a,k)}\ge0$ denote its eigenvalues. We define
the local $k$-neighbour anisotropy
\begin{equation}
	A_k^{(a)}
	=
	\frac{q_1^{(a,k)}-q_2^{(a,k)}}
	{q_1^{(a,k)}+q_2^{(a,k)}} ,
	\qquad 0\le A_k^{(a)}\le1 ,
	\label{eq:Ak}
\end{equation}
and its configuration average
\begin{equation}
	\mathcal A_k
	=
	\frac{1}{N}\sum_{a=1}^{N} A_k^{(a)} .
	\label{eq:AbarK}
\end{equation}
Under a rigid rotation, $Q^{(a,k)}$ transforms by orthogonal
conjugation, while under a global dilatation both eigenvalues acquire
the same multiplicative factor; translations leave the displacement
vectors unchanged. Hence $A_k^{(a)}$ and $\mathcal A_k$ are invariant
under translations, rotations, and global dilatations. A locally
isotropic neighbour distribution has $q_1\simeq q_2$ and
$A_k\simeq0$, whereas a strongly collinear local arrangement has
$q_1\gg q_2$ and $A_k\simeq1$.

Equation~\eqref{eq:Qk} uses displacements from the reference particle
$a$, rather than positions measured from the centroid of its neighbour
cloud. Thus $A_k^{(a)}$ quantifies the directional organization of the
neighbourhood as seen from particle $a$; it is not the covariance
anisotropy of the neighbours about their own centroid.

Our primary diagnostic uses $k=6$,
\begin{equation}
	A_6^{(a)}\equiv A_k^{(a)}\big|_{k=6},
	\qquad
	\Aani\equiv\mathcal A_k\big|_{k=6},
	\label{eq:A6def}
\end{equation}
motivated by the approximately sixfold local coordination of the
lower-variety planar configuration while retaining sensitivity to
local directional structure. No special status is assigned to $k=6$;
the dependence on neighbourhood size is examined explicitly below.

\begin{figure}[t]
	\centering
	\includegraphics[width=\columnwidth]{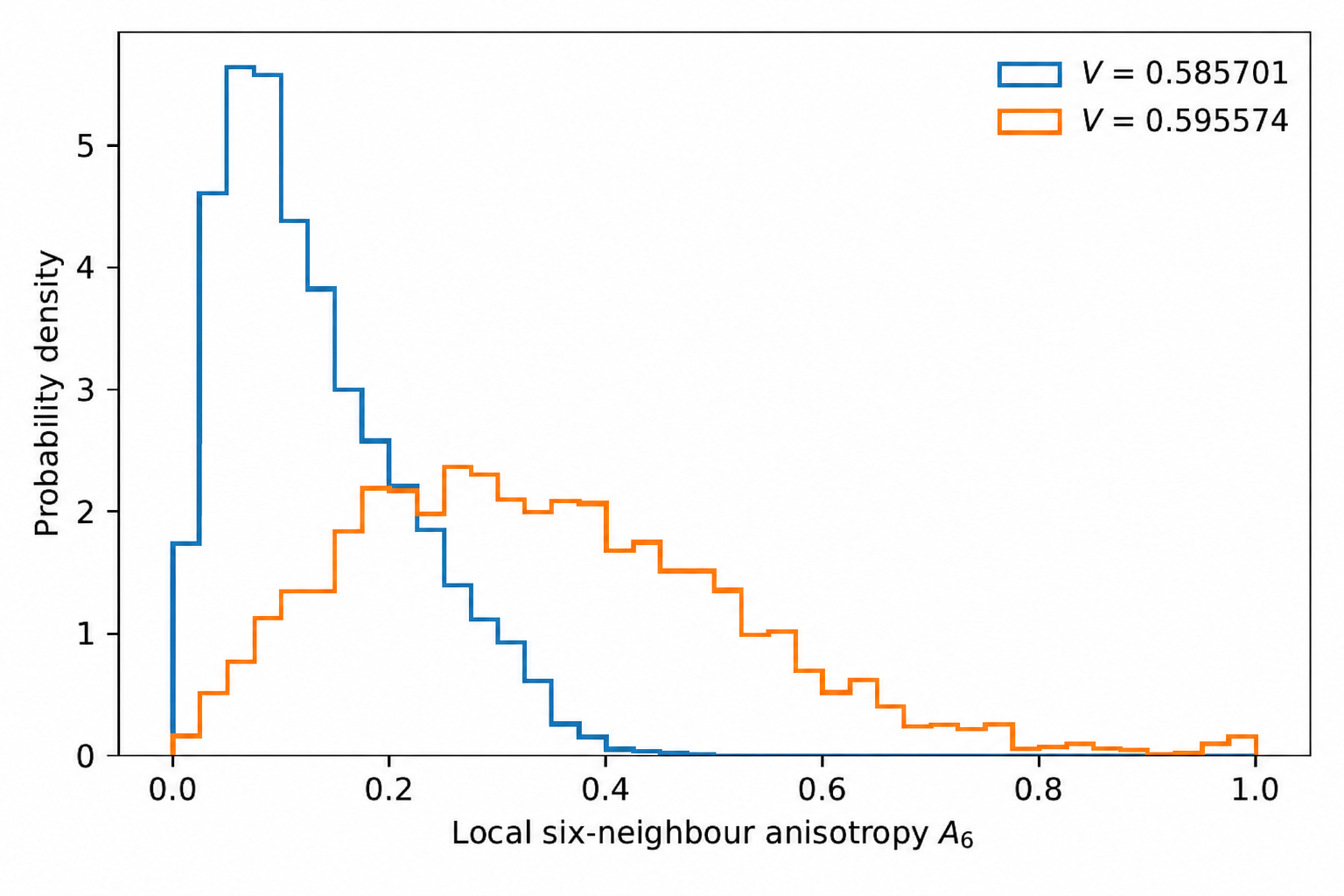}
	\caption{Probability density of the local six-neighbour anisotropy
		$A_6$ for the two $N=5000$ planar central configurations. The
		lower-variety configuration is concentrated at smaller anisotropy,
		whereas the higher-variety configuration exhibits a substantially
		broader distribution extending to larger $A_6$.}
	\label{fig:Ahist}
\end{figure}

\subsection{Numerical morphology results}

For the lower-variety configuration we obtain
\begin{equation}
	\Aani^{(0)}=0.133118,\qquad
	{\rm med}(A_6)^{(0)}=0.113344 ,
	\label{eq:A0}
\end{equation}
whereas the higher-variety configuration gives
\begin{equation}
	\Aani^{(1)}=0.341901,\qquad
	{\rm med}(A_6)^{(1)}=0.322639 .
	\label{eq:A3}
\end{equation}
Consequently,
\begin{equation}
	\frac{\Aani^{(1)}-\Aani^{(0)}}{\Aani^{(0)}}
	=1.568 ,
	\label{eq:Aincrease}
\end{equation}
so that the mean six-neighbour anisotropy of the higher-variety
configuration is $156.8\%$ larger. Equivalently,
\begin{equation}
	\frac{\Aani^{(1)}}{\Aani^{(0)}}=2.568 .
	\label{eq:factor}
\end{equation}
Combined with the $1.686\%$ variety separation established in
Eq.~\eqref{eq:Cdifference}, this shows that, for this pair of critical
shapes, a small difference in the global variety accompanies a much
larger relative difference in the chosen local morphology statistic.
This is a finite comparison between two configurations, not a
derivative or a dynamical response of $\mathcal A_6$ to $\variety$.

The corresponding probability densities are shown in
Fig.~\ref{fig:Ahist}. The lower-variety configuration is strongly
concentrated at small $A_6$, whereas the higher-variety configuration
has a much broader distribution extending to substantially larger local
anisotropy. Thus the difference between the two configurations is not
confined to their mean values but is visible across the full empirical
distributions.

The distributional change is also substantial. No particle in the
lower-variety configuration satisfies $A_6>0.5$, whereas
\begin{equation}
	\frac{N(A_6>0.5)}{N}=0.1808
	\label{eq:highA}
\end{equation}
for the higher-variety configuration. Thus $18.1\%$ of its particles
lie above this threshold. The value $A_6=0.5$ is used only as a
descriptive benchmark and is not assigned any special physical
significance. The empirical distributions have Kolmogorov--Smirnov
distance
\begin{equation}
	D_{\rm KS}=0.557 .
\end{equation}
We use $D_{\rm KS}$ solely to quantify their empirical separation.
Because local particle environments are spatially correlated, the
individual $A_6^{(a)}$ values are not treated as independent
observations, and no standard KS $p$-value is assigned.

\begin{figure}[t]
	\centering
	\includegraphics[width=\columnwidth]{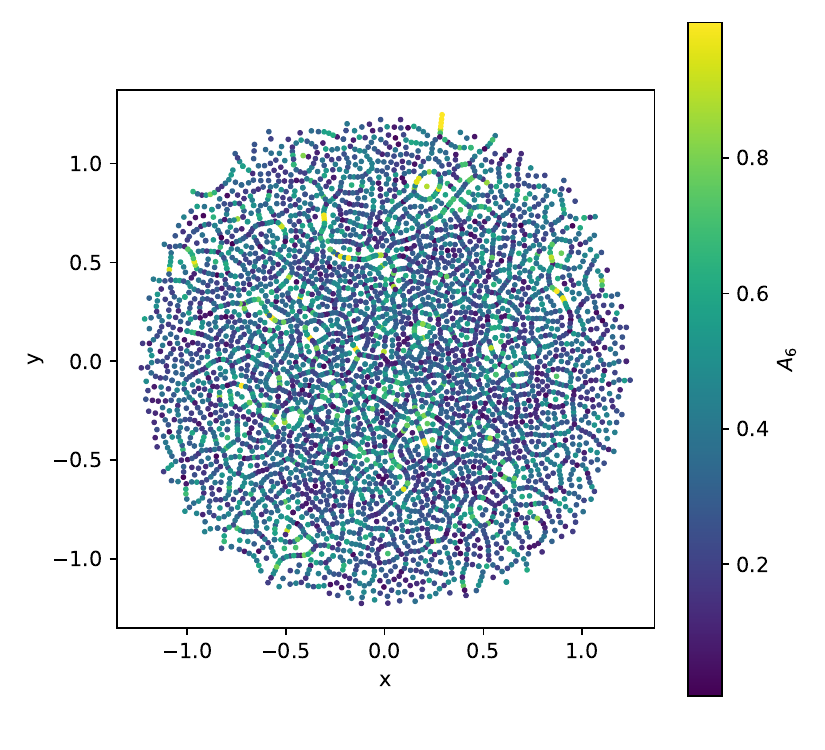}
	\caption{Higher-variety $N=5000$ planar central configuration with
		particles colored by their local six-neighbour anisotropy $A_6$.
		Larger values preferentially occur along the filamentary and curved
		structures, illustrating the geometrical information captured by the
		local diagnostic.}
	\label{fig:Amap}
\end{figure}

The spatial origin of the high-anisotropy tail is illustrated in
Fig.~\ref{fig:Amap}. Larger values of $A_6$ occur preferentially along
the filamentary and curved structures of the higher-variety
configuration, whereas more isotropic local environments carry smaller
values. Figure~\ref{fig:Amap} therefore connects the distributional
separation displayed in Fig.~\ref{fig:Ahist} with specific geometrical
features visible in the particle configuration itself.

\subsection{Dependence on neighbourhood size}

Because $\mathcal A_k$ is a local statistic, its numerical value depends
on the neighbourhood scale $k$. We therefore repeat the calculation for
$k=4,5,6,8,10,$ and $12$. The results are reported in
Table~\ref{tab:kcoarse-grained}.

As Table~\ref{tab:kcoarse-grained} shows, for every tested $k$ the
higher-variety configuration has the larger mean anisotropy. The
magnitude of the difference, however, is strongly scale dependent,
ranging from $8.8\%$ at $k=8$ to $161.8\%$ at $k=5$ and varying
nonmonotonically with $k$.
The present calculation therefore supports persistence of the
\emph{sign} of the anisotropy contrast over the tested neighbourhood
sizes, but not a scale-independent enhancement. Interpreting the
detailed $k$ dependence in terms of characteristic geometrical scales
would require a broader multiscale analysis.

\begin{table}[t]
	\caption{Mean local anisotropy as a function of the number $k$ of
		nearest neighbours. The final column gives the relative difference
		of the higher-variety configuration with respect to the lower-variety
		configuration.}
	\label{tab:kcoarse-grained}
	\begin{ruledtabular}
		\begin{tabular}{cccc}
			$k$ & $\mathcal A_k^{(0)}$ & $\mathcal A_k^{(1)}$
			& relative difference \\
			\hline
			4  & 0.2697 & 0.4966 & $84.1\%$ \\
			5  & 0.1568 & 0.4107 & $161.8\%$ \\
			6  & 0.1331 & 0.3419 & $156.8\%$ \\
			8  & 0.2424 & 0.2638 & $8.8\%$ \\
			10 & 0.1743 & 0.2175 & $24.8\%$ \\
			12 & 0.1147 & 0.1877 & $63.7\%$ \\
		\end{tabular}
	\end{ruledtabular}
\end{table}

\subsection{Sensitivity to very close particle pairs}

We finally test whether the measured anisotropy is dominated by very
close particle pairs. To preserve scale invariance, closeness is
defined through the dimensionless criterion
\begin{equation}
	\frac{r_{ab}}{\ell_{\rm rms}}<\epsilon .
	\label{eq:epsilon}
\end{equation}
For the higher-variety configuration, we construct a graph connecting
all pairs satisfying Eq.~\eqref{eq:epsilon}, with $\ell_{\rm rms}$
evaluated from the $N=5000$ configuration. Each connected
component containing more than one particle is replaced by its
geometrical centroid, and the $k=6$ anisotropy analysis is then
repeated on the resulting point set. This operation is used solely as
a geometrical robustness test: the modified point set is not assumed
to be a central configuration, and no variety is assigned to it.

At the reference value $\epsilon=10^{-4}$, the number of geometrical
points decreases from $5000$ to $4755$, yielding
\begin{align}
	\mathcal A_{6,{\rm cg}}^{(1)}&=0.316948,\\
	{\rm med}(A_6)_{\rm cg}^{(1)}&=0.300251,\\
	{\rm Prob}_{\rm cg}(A_6>0.5)&=0.125973 .
	\label{eq:coarse-grained}
\end{align}
The resulting mean anisotropy remains $138.1\%$ larger than
$\Aani^{(0)}$.

Varying $\epsilon$ from $10^{-5}$ to $2\times10^{-3}$ changes the
number of geometrical points only from $4759$ to $4754$ and gives
\begin{equation}
	0.31695\lesssim\mathcal A_{6,{\rm cg}}^{(1)}
	\lesssim0.31760 .
	\label{eq:epsrange}
\end{equation}
Over this tested interval, the modified-set mean anisotropy varies by
about $0.21\%$ relative to its minimum value. More importantly for the
present purpose, it remains well above the lower-variety value. The
principal anisotropy contrast is therefore not attributable solely to
the very closest particle pairs under this particular geometrical
robustness operation.

Table~\ref{tab:results} summarizes the original morphology statistics
together with the close-pair-modified point set. In particular, it
makes explicit that the reduction produced by the geometrical merging
operation does not erase the large separation from the lower-variety
configuration.

\begin{table}[t]
	\caption{Morphology statistics for the two planar CCs and for the
		geometrically modified point set obtained from the higher-variety CC
		at $\epsilon=10^{-4}$. The last column is included only as a robustness
		test; the modified point set is not assigned a central-configuration
		variety.}
	\label{tab:results}
	\begin{ruledtabular}
		\begin{tabular}{lccc}
			& lower-$\variety$ & higher-$\variety$ & modified point set \\
			\hline
			$\variety$ & 0.585701 & 0.595574 & --- \\
			$N$ & 5000 & 5000 & 4755 \\
			$\Aani$ & 0.1331 & 0.3419 & 0.3169 \\
			median $A_6$ & 0.1133 & 0.3226 & 0.3003 \\
			$P(A_6>0.5)$ & 0 & 0.1808 & 0.1260 \\
		\end{tabular}
	\end{ruledtabular}
\end{table}

\section{From the static shape-space landscape to the dynamical arrow}
\label{sec:bridge}

The numerical results obtained above permit a quantitative comparison
between the static structure of shape space and the dynamical
gravitational arrow, provided that the two are kept logically distinct.

The two planar CCs considered here are selected critical shapes of the
scale-invariant variety landscape. Their relevance to the BKM
construction is therefore indirect: the long-term branch-wise behavior
of $\variety$ characterizes the gravitational arrow, whereas the CCs
provide controlled static configurations at specified values of the
same global relational variable.

For the two configurations,
\begin{equation}
	\frac{\Delta\variety}{\variety^{(0)}}=1.686\%,
	\qquad
	\frac{\Delta\Aani}{\Aani^{(0)}}=156.8\% .
	\label{eq:staticfact}
\end{equation}
This is a finite comparison between two critical shapes, not an
evolution from one configuration to the other. It demonstrates that,
for this pair, nearby values of the global variety can coexist with
markedly different local geometrical organization.

The complementary evidence is displayed explicitly in the figures and
tables. Figure~\ref{fig:planarCCs} shows the global morphological
contrast between the two critical shapes; Fig.~\ref{fig:Ahist} shows
the corresponding separation of their local-anisotropy distributions;
and Fig.~\ref{fig:Amap} identifies where the large-$A_6$ values occur
within the higher-variety configuration. Table~\ref{tab:kcoarse-grained}
shows that the anisotropy ordering persists across all tested
neighbourhood sizes, while Table~\ref{tab:results} shows that the
principal contrast survives the close-pair robustness operation.
Taken together, these results show that the contrast is not attributable
solely to a particular neighbourhood choice or to a small number of
extreme local environments.

The result therefore establishes a property of the sampled static
shape-space landscape, but it does not establish a dynamical
morphology--variety relation. Two CCs cannot determine a universal or
monotonic relation between $\mathcal A_k$ and $\variety$, and generic
Newtonian histories are not required to pass through either
configuration.

The natural next step is dynamical. One should generate an ensemble of
generic solutions satisfying Eq.~\eqref{eq:sector}, identify the Janus
point of each history through $D=0$, and evaluate both $\variety$ and
the same morphology observable $\mathcal A_k$ along the two branches.
Since neither pointwise monotonicity of $\variety$ nor any monotonic
behavior of $\mathcal A_k$ is assumed, the relevant comparison should
be statistical rather than based on instantaneous derivatives. One
should determine whether the distribution of $\mathcal A_k$ changes
systematically with distance from the Janus region and how that
behavior is related to the contemporaneous value of $\variety$.

A reproducible statistical association between $\mathcal A_k$ and
$\variety$ along genuine histories would show that some of the
morphology--variety organization observed among critical shapes is also
sampled dynamically. Conversely, a weak, nonmonotonic, multivalued, or
highly scattered relation would indicate that the global variety
provides only a partial description of the local geometrical structure
encountered along the gravitational arrow.

A complementary static test is to enlarge the present CC sample. For
an ensemble of critical shapes $\{q_i\}$ spanning a controlled range of
variety, one may evaluate
\begin{equation}
	\left\{\variety_i,\mathcal A_{k,i}\right\}
	=
	\left\{\variety(q_i),\mathcal A_k(q_i)\right\},
	\label{eq:ensemble}
\end{equation}
using the same numerical prescription for every configuration. The
joint distribution $P(\mathcal A_k,\variety)$ would then reveal whether
the sampled critical-point landscape exhibits a systematic trend,
multiple morphological branches, a crossover, or substantial
degeneracy at fixed or nearly fixed variety.

These two extensions address different questions. A larger CC ensemble
tests how morphology is organized across the static critical-point
landscape, whereas genuine Janus-point trajectories test which aspects
of that organization, if any, are dynamically sampled. The central
question is therefore not whether the two CCs themselves form a
dynamical sequence, but whether the statistical relation between global
variety and local morphology across the critical-point landscape has a
counterpart along generic gravitational histories.

\section{Discussion and conclusions}
\label{sec:discussion}

The present analysis isolates a distinction that is central to a
relational description of gravitational structure. The variety
$\variety$ is a global, scale-invariant quantity constructed from all
pair separations, whereas $\mathcal A_k$ probes the directional
organization of local neighbourhoods. The two observables therefore
resolve different levels of relational information: $\variety$
characterizes global clustering contrast, while $\mathcal A_k$
quantifies local geometrical organization.

For the two planar $N=5000$ central configurations analyzed here,
\begin{equation}
	\begin{aligned}
		\left(\variety^{(0)},\mathcal A_6^{(0)}\right)
		&=(0.585701,0.1331),\\
		\left(\variety^{(1)},\mathcal A_6^{(1)}\right)
		&=(0.595574,0.3419).
	\end{aligned}
	\label{eq:summary}
\end{equation}
Thus, a $1.686\%$ difference in variety accompanies a $156.8\%$
difference in mean six-neighbour anisotropy. The distributional change
is likewise substantial: no particle in the lower-variety configuration
satisfies $A_6>0.5$, whereas the corresponding fraction in the
higher-variety configuration is $18.1\%$.

These complementary aspects of the result are visible in
Figs.~\ref{fig:planarCCs}--\ref{fig:Amap} and are summarized
numerically in Table~\ref{tab:results}: the configurations differ
visually, their $A_6$ distributions are strongly separated, and the
largest local anisotropies are spatially associated with the
filamentary and curved structures of the higher-variety configuration.
The contrast is therefore not reducible to a small number of extreme
local environments.

The comparison is robust in sign, but explicitly scale dependent.
Table~\ref{tab:kcoarse-grained} shows that the higher-variety CC is
more anisotropic for every tested neighbourhood size,
$4\le k\le12$, although the magnitude of the difference varies
appreciably with $k$.
This scale dependence is itself informative: local morphology is
intrinsically multiscale, whereas $\variety$ compresses global
relational information into a single scalar. The close-pair robustness
test leads to the same qualitative conclusion. At the reference
threshold, the modified point set retains a mean anisotropy $138.1\%$
above that of the lower-variety configuration, with only about $0.21\%$
variation over the tested threshold range.
The corresponding before-and-after statistics are collected in
Table~\ref{tab:results}.

The central result is therefore deliberately limited but conceptually
sharp: for this pair of critical shapes, nearby values of the global
variety coexist with markedly different local geometrical organization.
Variety does not, by itself, uniquely specify morphology. This
conclusion does not diminish its role in the BKM construction. Rather,
it clarifies that a global scalar whose long-term branch-wise behavior
characterizes the gravitational arrow need not constitute a complete
descriptor of the structural state of the system.

Central configurations are particularly useful for exposing this
distinction because they are controlled critical points of the
scale-invariant shape-space landscape. They are not hidden attractors
of generic Newtonian dynamics, and the two configurations studied here
must not be interpreted as successive stages of a single history.
Their value is instead diagnostic: they show that substantial
morphological freedom can exist within a narrow interval of the same
global relational variable.

This observation opens two complementary directions. The first is
static: a substantially larger ensemble of central configurations
would determine the joint distribution
$P(\mathcal A_k,\variety)$ across the critical-point landscape and
reveal whether the present pair lies on a smooth morphological trend,
on distinct structural branches, near a crossover, or within a broad
degeneracy at fixed variety. The second is dynamical: ensembles of
generic Janus-point histories should be used to measure the same
morphology observable along both branches and determine whether the
distribution of $\mathcal A_k$ is statistically organized by
contemporaneous variety, by distance from the Janus region, or by
additional relational variables.

These questions define a natural bridge between the geometry of shape
space and the dynamics of the gravitational arrow. If a reproducible
morphology--variety relation is found along genuine histories, the
static CC landscape would acquire direct dynamical significance as a
map of structural organization accessible to the evolving system. If,
instead, morphology remains strongly multivalued at fixed variety,
that result would be equally important: it would show that the
gravitational arrow is intrinsically higher-dimensional and that a
single global scalar captures its orientation without exhausting its
structural content.

The broader lesson is therefore not that variety should be replaced,
but that it may form the first level of a hierarchy of relational
descriptors. Global variety quantifies clustering contrast; local
anisotropy resolves directional organization; further observables may
be required to capture connectivity, filamentarity, void structure, or
other aspects of emergent geometry. Establishing how such descriptors
interrelate, and which of them acquire systematic branch-wise behavior
along generic Newtonian histories, offers a concrete route toward a
more complete theory of structural emergence in self-gravitating
systems.

In this sense, the present result should be viewed as a first
quantitative step from an arrow characterized by global relational
structure toward an arrow resolved in morphology. The decisive question
for future work is no longer simply how global gravitational clustering
develops away from the Janus region, but how the geometry of structure
itself is organized, differentiated, and dynamically sampled along
that evolution.

\begin{acknowledgments}
	F.S.N.L. acknowledges support from Funda\c{c}\~ao para a Ci\^encia e a Tecnologia through contract CEECINST/00032/2018 and grants UID/04434/2025 and PTDC/FIS-AST/0054/20. M.I.R.L. acknowledges the Instituto de Astrof\'isica e Ci\^encias do Espa\c{c}o for computational resources. J.B. thanks Tim Koslowski for many helpful discussions. 
\end{acknowledgments}

\end{document}